\documentclass[proceedings]{JHEP3} 

\usepackage{epsfig,multicol}			

\title{{\bf Monodromy, Duality and Integrability of Two Dimensional
    String Effective Action}}

\author{\speaker{Ashok Das}$^\dagger$,
J. Maharana$^\ddagger$ and  A. Melikyan$^\dagger$ }
\author{{$^\dagger$}                                           
{\it Department of Physics and Astronomy\\University of Rochester, Rochester, NY 14627-0171, USA.}\\
      E-mail: \email{das@pas.rochester.edu},
      \email{arsen@pas.rochester.edu}}
\author{{$^\ddagger$}
\it {Institute of Physics\\Bhubaneswar -751005, India}\\
      E-mail: \email{maharana@iopb.res.in}}

\conference{Workshop on Integrable Field Theories, Solitons and Duality.}

\abstract{In this talk, we show how the monodromy matrix, ${\widehat{\cal M}}$,
can be  constructed for the  
two dimensional tree level string effective action. The pole structure
of ${\widehat{\cal M}}$ is derived using its factorizability
property. It is shown that the monodromy matrix transforms non-trivially
under the non-compact T-duality group, which leaves the effective action
invariant and this can be used to construct the monodromy matrix for
more complicated backgrounds starting from simpler ones. We construct,
explicitly,  ${\widehat{\cal M}}$ for
the exactly solvable Nappi-Witten model, both when $B=0$ and $B\neq
0$,  where these ideas can be
directly checked.}

\begin{document}

\vspace{.4in}

\section{Introduction}
The field theories in two space-time dimensions have attracted
considerable attention over the past few decades. They possess a
variety of interesting features. Some of these field theories,
capture several salient characteristics of four dimensional
theories and, therefore, such two dimensional models are used as
theoretical laboratories. Moreover, the nonperturbative properties
of field theories are much simpler to study in two dimensional
models. There are  classes of two dimensional theories which are
endowed with a rich symmetry structure: integrable models
\cite{das} and conformal field theories \cite{abd,cft} belong to
this special category among others. Under special circumstances,
in the presence of isometries, a four dimensional theory may also
be described by an effective
 two dimensional theory.

The string theories are abundantly rich in their symmetry
content.
 The tree level string
 effective action, dimensionally reduced to lower dimensions, is known to
possess enlarged symmetries \cite{gpr,as,jm}.
 Let us consider, toroidal compactification
of heterotic string on a $d$-dimensional torus, from
$10$-dimensional space-time to $10-d$ dimensions. The reduced
theory is known to be invariant under the non-compact T-duality
group, $O(d, d+16)$. For the case $d=6$,
 namely, in the case of reduction to
four space-time dimensions, the field strength of the two form
anti-symmetric
 tensor can be traded for the pseudoscalar axion. Furthermore, the dilaton
and axion can be combined to parameterize the coset
${SL(2,R)}\over{U(1)}$. Thus, the four dimensional theory possesses
the T-duality as well
 as the S-duality group of symmetries. When the string effective
 action is  reduced to two space-time
dimensions we encounter an enhancement in symmetry as has been
studied
 by  several
authors \cite{ib,jmi,jhs,sen}. The string effective action
describes supergravity
 theories and
the integrability properties of such theories have been
investigated in the recent past \cite{nic2,nic3}. It is worth
while to mention that higher dimensional Einstein theory,
dimensionally reduced to effective two dimensional theories, has
been studied in the past \cite{bm}.
 One of the approaches is to derive the monodromy matrix
which encodes some of the essential features of integrable field
theories. An effective two dimensional action naturally appears
when one
 considers some  aspects
of black hole physics, colliding plane waves as well as
 special types of cosmological models.
Recently, we have shown that the two dimensional string effective
action has connection with integrable systems from  a new
perspective in the sense that one can construct the monodromy
matrix for such theories with well defined prescriptions
\cite{dmm1}. It was shown, while investigating the collision of
plane fronted stringy waves, that the monodromy matrix can be
constructed explicitly for a given set of background
configurations \cite{dmm}.
 Subsequently, we
were able to give the procedures for deriving the monodromy
matrix under general settings. It is worth while to mention that
some of the interesting aspects of black hole physics can be
described by an effective two dimensional theory \cite{cghsb}. Moreover, there
is an intimate relation between colliding plane waves and the description
of four dimensional space-time with two commuting Killing
 vectors. 

Chandrasekhar and Xanthopoulos \cite{cx},
 in their  seminal work, have shown that
a violently time dependent space-time with a pair of Killing
vectors provides a description of plane colliding gravitational
waves. Furthermore, Ferrari, Ibanez and Bruni \cite{fib} have
demonstrated that the colliding plane wave metric can be
identified,  locally, to be isometric to the interior of a
Schwarzschild metric. In another important step, Yurtsever
\cite{yu}
 constructed the transformation which
 provides a connection between a metric describing colliding waves
and one corresponding to the Schwarzschild black hole. As is
well known, from the study of colliding plane gravitational waves
\cite{grf} corresponding to massless states of strings,
there is a curvature singularity in the future. Recently, the
appearance of the future singularity, has found an alternative
description in the context of Pre-Big Bang (PBB) scenario \cite{fkv,bv}.
The incoming plane waves are to be identified as the initial
stringy vacuua of the Universe, which collide and lead to the
creation of the Universe. Indeed, the solutions correspond to
Kasner type metric and the exponents fulfill the requirements of
PBB conditions. A very important fact, in this context, is that one
 starts  from a
four dimensional effective action; however, the physical process
is effectively described  by a two dimensional theory. 

When we
focus our attention to address these problems in the frame work of
 string theory, it is essential to keep in mind the special symmetries, such
as dualities, which are an integral part of the stringy symmetries. We
have investigated \cite{dmm1} the behavior of the monodromy
matrix under T-duality transformation of the backgrounds under a
general setting when the two dimensional action is derived from a
$D$-dimensional string effective action through compactification
on a $d$-dimensional torus, $T^d$. It was shown that the monodromy
matrix transforms non-trivially under the duality group $O(d,d)$.
Therefore, it opens up the possibility of studying integrable
systems which might appear in the context of string theories. As
an example, we considered the Nappi-Witten \cite{nw} model, which is
a solution to an exact conformal field theory described by a WZW
model. We first obtained the monodromy matrix for the case when the 
2-form anti-symmetric tensor is set to zero. It is also well known
that an anti-symmetric tensor background can be generated through an
$O(2,2)$ transformation from the initial backgrounds \cite{gmv}.
 We can construct the monodromy
matrix for the new set of backgrounds using our prescriptions. On
the other hand, the new monodromy matrix can be constructed
directly by utilizing the transformation rules discovered by us.
Indeed, we explicitly, demonstrated that the monodromy matrices
obtained via the two different routes are identical. It is obvious
from the preceding discussions that the symmetries of the
effective action play an important role in the construction of the
monodromy matrix and its transformation properties under those
symmetry transformations. Therefore, it is natural to expect
intimate connections between the integrability properties of the
two dimensional theory and the full stringy symmetry groups of T
and S dualities.

In this lecture, we review the works reported in
\cite{dmm1,dmm,dmm2}. We provide
prescriptions for the construction of the monodromy matrix,
$\widehat{\cal M}$, for the string theoretic two dimensional
effective action. We present the pole structure of the monodromy
matrix from general arguments. The duality transformation
properties of $\widehat{\cal M}$ follow from the definition and
construction of this matrix. For example, if the action respects
T-duality, then one can derive how $\widehat{\cal M}$ transforms
under the group, whereas, if the underlying symmetry corresponds
to S-duality an appropriate transformation rule for monodromy
matrix can also be derived.
Our paper is organized as follows: In Section {\bf II}, we
recapitulate the form of the two dimensional action obtained by
toroidal compactification from higher dimensions. Then, we present
the equations of motion. 
A key ingredient, in the derivation of the monodromy matrix, is the
coset  space reformulation of the reduced
action. We devote Section {\bf III} to the construction of the
monodromy  matrix for the problem under consideration. The
transformation rules for $\widehat{\cal M}$, under a T-duality
transformation,  are derived once the
matrix is constructed. An interesting observation is that the
expression for $\widehat{\cal M}$ already captures the stringy
symmetry in an elegant manner. In this section, we also present
explicit forms of $\widehat{\cal M}$ for simple background
configurations which still preserve some of the general features.
We present an illustrative example in Section {\bf IV}, namely, the
structure of the Nappi-Witten model in the present context is analyzed
in detail. We present a
brief conclusion in Section {\bf V}.

\section{Two dimensional effective action}

In this section, we will briefly recapitulate the form of the string
effective action in two dimensions, which will form the basis for all our
subsequent discussions. Let us consider, for simplicity, the tree level
string effective action in $D$-dimensions consisting of the graviton, the
dilaton and the anti-symmetric tensor field,

\begin{equation}
\hat{S}=\int d^{D}x\sqrt{-\hat{G}}e^{-\hat{\phi }}\left( R_{\hat{G}}+\left(
\hat{\partial }\hat{\phi }\right) ^{2}-\frac{1}{ 12}\hat{H}_{\hat{\mu }\hat{
\nu }\hat{\rho }}\hat{H}^{\hat{\mu }\hat{\nu }\hat{\rho }}\right) ,
\label{action1}
\end{equation}
Here, $\hat{G}_{\hat{\mu }\hat{\nu }}$ is the $D$-dimensional metric in the
string frame with signature $(-,+,\cdots, +)$, and $\hat{G}=\det \hat{G}_{%
\hat{\mu}\hat{\nu }}$. $\hat{R}_{\hat{G}}$ is the scalar curvature, $\hat{
\phi}$ is the dilaton and $\hat{H}_{\hat{\mu }\hat{\nu }\hat{\rho }
}=\partial_{\hat{\mu }}\hat{B}_{\hat{\nu } \hat{\rho }}+\partial_{\hat{\rho }
}\hat{B}_{\hat{\mu } \hat{\nu }}+\partial_{\hat{\nu }}\hat{B}_{\hat{\rho }
\hat{\mu }}$ is the field strength for the second-rank anti-symmetric tensor
field, $\hat{B}_{\hat{\mu }\hat{\nu }}.$

If we compactify this action on a $d$-dimensional torus, $T^{d}$, where $d =
D-2$, then, the resulting dimensionally reduced action would describe the
two dimensional string effective action, which has the form \cite{jmj,hasan},
\begin{equation}
S=\int dx^{0}dx^{1}\sqrt{-g}e^{-\bar{\phi}}\left( R+\left( \partial \bar{%
\phi }\right) ^{2}+\frac{1}{8}\,{\rm Tr}\,\left(\partial_{\alpha
}M^{-1}\partial^{\alpha}M\right) \right)  \label{action2}
\end{equation}
Here $\alpha ,\beta =0,1$ are the two-dimensional space-time indices, $
g_{\alpha\beta }$ is the two-dimensional space-time metric with $g=\det
g_{\alpha \beta }$. $R$ is the corresponding two dimensional scalar
curvature, while the shifted dilaton is defined as
\begin{equation}
\bar{\phi}=\phi -\frac{1}{2}\log \det G_{ij}  \label{ShiftDilaton}
\end{equation}
where $G_{ij}$ is the metric in the internal space, corresponding to the 
toroidally compactified coordinates
 $x^{i}$, $i,j=2,3,...,D-1$ $(d=D-2)$. Finally, $M$ is
a $2d\times 2d$ symmetric matrix of the form
\begin{equation}
M=\left(
\begin{array}{cc}
G^{-1} & -G^{-1}B \\
BG^{-1} & G-BG^{-1}B%
\end{array}
\right)  \label{M-matrix}
\end{equation}
where $B$ represents the moduli coming from the dimensional reduction of the
$\hat{B}$-field in $D$-dimensions. In general, there will be additional
terms in (\ref{action2}) associated with $d$ Abelian gauge fields
arising from the metric, $\hat{G}_{\hat{\mu}\hat{\nu}}$, and another set of $
d$ Abelian gauge fields coming from the anti-symmetric tensor, $\hat{B}_{%
\hat{\mu}\hat{\nu}}$, as a result of dimensional reduction \cite{jmj} .
Furthermore, there would also have been terms involving the field strength
of the two dimensional tensor field, $B_{\alpha \beta }$. Since we are in
two space-time dimensions, we have dropped the gauge field terms and, in the
same spirit, have not kept the field strength of $B_{\alpha \beta}$, which
can always be removed, if it depends only on the coordinates $x^{0}$ and $
x^{1}$. Later, we will comment on the gauge fields, which assume a
significant role, when Abelian gauge fields are present in the original
string effective action (\ref{action1}).

The matrix $M$ corresponds to a symmetric representation of the group $
O(d,d) $ and the dimensionally reduced action in (\ref{action2}) is
invariant under the global $O(d,d)$ transformations
\begin{eqnarray}
g_{\alpha \beta } &\rightarrow &g_{\alpha \beta },\quad\bar{\phi } \rightarrow
\bar{\phi }  \label{trans1} \\
&&  \nonumber \\
M &\rightarrow &\Omega ^{T}M\Omega  \label{trans2}
\end{eqnarray}
where $\Omega \in O(d,d)$ is the global transformation matrix, which
preserves the $O(d,d)$ metric

\begin{equation}
\eta =\left(
\begin{array}{cc}
0 & {\bf 1}_{d} \\
{\bf 1}_{d} & 0%
\end{array}
\right)  \label{eta}
\end{equation}
with ${\bf 1}_{d}$ representing the identity matrix in $d$
dimensions. Namely, $\Omega$ satisfies $\Omega^{T}\eta\Omega = \eta$.

The equations of motion for the different fields follow from the
dimensionally reduced effective action (\ref{action2}). For example, varying
the effective action (\ref{action2}) with respect to the shifted dilaton, $
\bar{\phi}$, and the metric, $g_{\alpha\beta}$, leads respectively to

\begin{eqnarray}
R+2g^{\alpha \beta }D_{\alpha }D_{\beta }\bar{\phi }-g^{\alpha \beta
}\partial _{\alpha }\bar{\phi }\partial _{\beta }\bar{\phi }+\frac{ 1}{8}
g^{\alpha \beta }Tr\left( \partial _{\alpha }M^{-1}\partial _{\beta
}M\right) &=&0  \label{eq1} \\
&&  \nonumber \\
R_{\alpha \beta }+D_{\alpha }D_{\beta }\bar{\phi }+\frac{1}{8}Tr\left(
\partial _{\alpha }M^{-1}\partial _{\beta }M\right) &=&0  \label{eq2}
\end{eqnarray}
It follows from these that
\begin{equation}
D_{\alpha}D^{\alpha} e^{-\bar{\phi}} = 0  \label{eq4aa}
\end{equation}
The variation of the effective action with respect to $M$ needs some care
since $M$ is a symmetric $O(d,d)$ matrix satisfying $M\eta M=\eta$. A simple
method, for example, would involve adding the constraint to the effective
action through a Lagrange multiplier. In any case, since there is no
potential term in the effective action involving the matrix $M$, the
Euler-Lagrange equation of motion following from the variation of the
action with respect to $M$ has the form of a conservation law
\begin{equation}
\partial _{\alpha }\left( e^{-\bar{\phi }}\sqrt{-g}g^{\alpha \beta
}M^{-1}\partial _{\beta }M\right) = 0  \label{eq3}
\end{equation}

We can further simplify these equations by working in the light-cone
coordinates
\begin{equation}
x^{+}=\frac{1}{\sqrt{2}}(x^{0}+x^{1}),\qquad x^{-}=\frac{1}{\sqrt{2}}
(x^{0}-x^{1})  \label{lightcone}
\end{equation}
and choosing the conformal gauge for the two dimensional metric, namely,
\begin{equation}
g_{\alpha \beta }=e^{F(x^{+},x^{-})}\eta _{\alpha \beta }  \label{conformal}
\end{equation}
In this case, eqs. (\ref{eq4aa}) and (\ref{eq3}) take respectively the forms
\begin{eqnarray}
\partial _{+}\partial _{-}e^{-\bar{\phi}} &=&0  \label{dilaton} \\
\partial _{+}\left( e^{-\bar{\phi}}M^{-1}\partial _{-}M\right) +\partial
_{-}\left( e^{-\bar{\phi}}M^{-1}\partial _{+}M\right) &=&0  \label{sigma}
\end{eqnarray}
while eqs. (\ref{eq1}) and (\ref{eq2}) can be written explicitly as
\begin{eqnarray}
\partial _{+}^{2}\bar{\phi}-\partial _{+}F\partial _{+}\bar{\phi}+{\frac{1}{
8 }}\,{\rm Tr}\,\left( \partial _{+}M^{-1}\partial _{+}M\right) &=&0
\nonumber \\
\partial _{+}\partial _{-}\bar{\phi}-\partial _{+}\partial _{-}F+{\frac{1}{8}
}\,{\rm Tr}\,\left( \partial _{+}M^{-1}\partial _{-}M\right) &=&0  \nonumber
\\
\partial _{-}^{2}\bar{\phi}-\partial _{-}F\partial _{-}\bar{\phi}+{\frac{1}{
8 }}\,{\rm Tr}\,\left( \partial _{-}M^{-1}\partial _{-}M\right) &=&0
\end{eqnarray}

It is well known \cite{jmj} that the moduli appearing in the definition of
the $M$ matrix (see eq. (\ref{M-matrix})) parameterize the coset ${\frac{O(d,d)
}{O(d)\times O(d)}}$. Correspondingly, it is convenient to introduce a
triangular matrix $V\in {\frac{O(d,d)}{O(d)\times O(d)}}$, of the form
\begin{equation}
V=\left(
\begin{array}{cc}
E^{-1} & 0 \\
BE^{-1} & E^{T}
\end{array}
\right)  \label{V-Matrix}
\end{equation}
such that $M=VV^{T}$. Here, $E$ is the vielbein in the internal space so
that $(E^{T}E)_{ij}=G_{ij}$. Under a combined global $O(d,d)$ and a local $
O(d)\times O(d)$ transformation
\begin{equation}
V\longrightarrow \Omega ^{T}Vh(x)
\end{equation}
where $\Omega \in O(d,d)$ and $h(x)\in O(d)\times O(d)$,
\begin{equation}
M=VV^{T}\longrightarrow \Omega ^{T}M\Omega
\end{equation}
Namely, the $M$ matrix is sensitive only to a global $O(d,d)$ rotation.

From the matrix $V$, we can construct the current $V^{-1}\partial _{\alpha
}V $ , which belongs to the Lie algebra of $O(d,d)$ and can be decomposed as
\begin{equation}
V^{-1}\partial _{\alpha }V=P_{\alpha }+Q_{\alpha }
\end{equation}
Here, $Q_{\alpha }$ belongs to the Lie algebra of the maximally compact
subgroup $O(d)\times O(d)$ and $P_{\alpha }$ belongs to the complement.
Furthermore, it follows from the symmetric space automorphism property of
the coset ${\frac{O(d,d)}{O(d)\times O(d)}}$ that $P_{\alpha }^{T}=P_{\alpha
},Q_{\alpha }^{T}=-Q_{\alpha }$ so that we can identify
\begin{eqnarray}
P_{\alpha } &=&{\frac{1}{2}}\left( V^{-1}\partial _{\alpha
}V+(V^{-1}\partial _{\alpha }V)^{T}\right)  \nonumber \\
Q_{\alpha } &=&{\frac{1}{2}}\left( V^{-1}\partial _{\alpha
}V-(V^{-1}\partial _{\alpha }V)^{T}\right)  \label{current}
\end{eqnarray}

It is now straightforward to check that
\begin{equation}
{\rm Tr}\,\left( \partial _{\alpha }M^{-1}\partial_{\beta}M\right) = - 4
{\rm Tr}\,\left( P_{\alpha }P_{\beta }\right)  \label{connection}
\end{equation}
Furthermore, under a global $O(d,d)$ rotation, the currents in
(\ref{current})  are invariant, while under a local $O(d)\times O(d)$
transformation, $ V\longrightarrow Vh(x)$,
\begin{equation}
P_{\alpha }\longrightarrow h^{-1}(x)P_{\alpha }h(x),\qquad Q_{\alpha
}\longrightarrow h^{-1}(x)Q_{\alpha }h(x)+h^{-1}(x)\partial _{\alpha }h(x)
\label{PQ-transform}
\end{equation}
Namely, under a local $O(d)\times O(d)$ transformation, $Q_{\alpha}$
transforms like a gauge field, while $P_{\alpha}$ transforms as belonging to
the adjoint representation. It is clear, therefore, that (\ref{connection})
is invariant under the global $O(d,d)$ as well as the local $O(d)\times O(d)$
transformations. Consequently, the action in (\ref{action2}) is also
invariant under the local $O(d)\times O(d)$ transformations.

This brings out, naturally, the connection between the system under study
and two dimensional integrable systems. First, let us note that, in the
absence of gravity and the dilaton (namely, if $g_{\alpha\beta}=\eta_{\alpha
\beta},\bar{\phi}=0$), the action in (\ref{action2}) simply corresponds to a
flat space sigma model defined over the coset ${\frac{O(d,d)}{O(d)\times
O(d) }}$, which can be analyzed through a zero curvature condition with a
constant spectral parameter (to be discussed in more detail in the next
section). In the presence of gravity as well as the dilaton, we can
eliminate the dilaton from the action (\ref{action2}) by choosing the
particular conformal gauge $g_{\alpha\beta} = e^{\bar{\phi}}
\eta_{\alpha\beta}$. In this case, the action will describe a sigma model,
defined over the coset ${\frac{O(d,d)}{O(d)\times O(d)}}$, coupled to
gravity. As we will show in the next section, this system can also be
analyzed through a zero curvature condition much like the flat space case,
although consistency requires the spectral parameter, in this case, to be
space-time dependent.

So far, we have only discussed the two dimensional string effective action
starting from the $D$-dimensional action in (\ref{action1}) involving the
graviton, the dilaton and the anti-symmetric tensor field. However, the
action in (\ref{action1}) can be generalized by adding $n$ Abelian gauge
fields, with the additional action of the form (such terms naturally arise
in heterotic string theory)
\begin{equation}
\hat{S}_{\hat{A}}=-\frac{1}{4}\int d^{D}x\sqrt{-\hat{G}}e^{-\hat{\phi}}\left(
\hat{g}^{\hat{\mu}\hat{\rho}}\hat{g}^{\hat{\nu}\hat{\lambda}}\delta _{IJ}
\hat{F}_{\hat{\mu}\hat{\nu}}^{I}\hat{F}_{\hat{\rho}\hat{\lambda}}^{J}\right)
\label{addterm}
\end{equation}
where $I,J=1,2,...,n$ and
\begin{equation}
\hat{F}_{\hat{\mu}\hat{\nu}}^{I}=\partial _{\hat{\mu}}\hat{A}_{\hat{\nu}
}^{I}-\partial _{\hat{\nu}}\hat{A}_{\hat{\mu}}^{I}  \label{F=dA}
\end{equation}
This action can also be dimensionally reduced \cite{jmj}
 to two dimensions and the
resulting effective action takes the form
\begin{equation}
S_{A}=-\frac{1}{4}\int dx^{0}dx^{1}\sqrt{-g}e^{-\bar{\phi}}\left( F_{\alpha
\beta }^{I}F^{I\alpha \beta }+2F_{\alpha j}^{I}F^{I\alpha j}\right)
\label{reducedA}
\end{equation}
where we have defined
\begin{eqnarray}
a_{i}^{I} &=&\hat{A}_{i}^{I}  \nonumber \\
A_{\alpha }^{(1)\,I} &=&\hat{G}_{\alpha }^{I}  \nonumber \\
A_{\alpha }^{(3)\,I} &=&\hat{A}_{\alpha }^{I}-a_{j}^{I}A_{\alpha }^{(1)j}
\nonumber \\
F_{\alpha \beta }^{(1)\,i} &=&\partial _{\alpha }A_{\beta
}^{(1)\,i}-\partial _{\beta }A_{\alpha }^{(1)\,i}  \label{defin1} \\
F_{\alpha \beta }^{(3)\,I} &=&\partial _{\alpha }A_{\beta }^{(3)I}-\partial
_{\beta }A_{\alpha }^{(3)I}  \nonumber \\
F_{\alpha \beta }^{I} &=&F_{\alpha \beta }^{(3)I}+F_{\alpha \beta
}^{(1)\,i}a_{i}^{I}  \nonumber \\
F_{\alpha i}^{I} &=&\partial _{\alpha }a_{i}^{I}  \nonumber
\end{eqnarray}
In the presence of the Abelian gauge fields, the field strength, $H$,
associated with the second rank anti-symmetric tensor field, $B$, needs to
be redefined for gauge invariance as
\begin{eqnarray}
H_{\alpha ij} &=&\partial _{\alpha }B_{ij}+\frac{1}{2}\left(
a_{i}^{I}\partial _{\alpha }a_{j}^{I}-a_{j}^{I}\partial _{\alpha
}a_{i}^{I}\right)  \nonumber \\
&&  \label{def2} \\
H_{\alpha \beta i} &=&-C_{ij}F_{\alpha \beta }^{(1)\,j}+F_{\alpha \beta
i}^{(2)}-a_{i}^{I}F_{\alpha \beta }^{(3)I}  \nonumber \\
&&  \nonumber \\
H_{\alpha \beta \gamma } &=&\partial _{\alpha }B_{\beta \gamma }-\frac{1}{2}
{\cal A}_{\alpha }^{r}\eta _{rs}{\cal F}_{\beta \gamma }^{s}+{\rm 
cyc.\,perms.}  \nonumber
\end{eqnarray}
where ${\cal A}_{\alpha }^{r}=(A_{\alpha }^{(1)\,i},A_{\alpha
i}^{(2)},A_{\alpha }^{(3)I})$, ${\cal F}_{\alpha \beta }^{r}=\partial
_{\alpha }{\cal A}_{\beta }^{r}-\partial _{\beta }{\cal A}_{\alpha }^{r}$
and
\begin{eqnarray}
A_{\alpha i}^{(2)} &=&\hat{B}_{\alpha i}+B_{ij}A_{\alpha }^{(1)\,j}+\frac{1}{
2}a_{i}^{I}A_{\alpha }^{(3)I}  \nonumber \\
F_{\alpha \beta \,i}^{(2)} &=&\partial _{\alpha }A_{\beta i}^{(2)}-\partial
_{\beta }A_{\alpha i}^{(2)}  \nonumber \\
C_{ij} &=&\frac{1}{2}a_{i}^{I}a_{j}^{I}+B_{ij}  \label{def3}
\end{eqnarray}

Once again, it is easy to see that, in two space-time dimensions, the field
strength $H_{\alpha \beta \gamma }$ can be set to zero. Furthermore, keeping
all other terms, the complete two dimensional string effective action can be
shown to have the same form as in (\ref{action2}) with
\begin{equation}
M=\left(
\begin{array}{ccc}
G^{-1} & -G^{-1}C & -G^{-1}a^{T} \\
-CG^{-1} & G+C^{T}G^{-1}C & C^{T}G^{-1}a^{T}+a^{T} \\
-aG^{-1} & aG^{-1}C+a & 1+aG^{-1}a^{T}
\end{array}
\right)  \label{M2}
\end{equation}
In this case, $M$ is a symmetric $d\times (d+n)$ matrix ($d=D-2$) belonging
to $O(d,d+n)$. Under a $O(d,d+n)$ transformation
\begin{equation}
M\longrightarrow \Omega ^{T}M\Omega
\end{equation}
where the parameter of transformation $\Omega \in O(d,d+n)$ satisfying $
\Omega ^{T}\eta \Omega =\eta $, where
\begin{equation}
\eta =\left(
\begin{array}{ccc}
0 & {\bf 1}_{d} & 0 \\
{\bf 1}_{d} & 0 & 0 \\
0 & 0 & {\bf 1}_{n}
\end{array}
\right)  \label{eta2}
\end{equation}
represents the metric for $O(d,d+n)$. As in the earlier case, it is more
convenient to introduce a matrix $V\in {\frac{O(d,d+n)}{O(d)\times O(d+n)}}$
of the form
\begin{equation}
V=\left(
\begin{array}{ccc}
E^{-1T} & 0 & 0 \\
-C^{T}E^{-1T} & E^{T} & a^{T} \\
-aE^{-1T} & 0 & 1
\end{array}
\right)  \label{V2}
\end{equation}
such that $M=VV^{T}$. As before, under a combined global $O(d,d+n)$ and a
local $O(d)\times O(d+n)$ transformation
\begin{equation}
V\longrightarrow \Omega ^{T}Vh(x)
\end{equation}
where $\Omega \in O(d,d+n)$ and $h(x)\in O(d)\times O(d+n)$. However, the
matrix $M$ is not sensitive to the local $O(d)\times O(d+n)$
transformations. We can now define the current $V^{-1}\partial _{\alpha }V$
which belongs to the Lie algebra of $O(d,d+n)$ and which can be decomposed
as
\begin{equation}
V^{-1}\partial _{\alpha }V=P_{\alpha }+Q_{\alpha }  \label{decomp2}
\end{equation}
In the present case, $Q_{\alpha }$ belongs to the Lie algebra of the maximal
compact subgroup $O(d)\times O(d+n)$, while $P_{\alpha }$ belongs to the
complement. Under a global $O(d,d+n)$ transformation, $P_{\alpha }$ and $
Q_{\alpha }$ are invariant, while under a local $O(d)\times O(d+n)$
transformation, $V\rightarrow Vh(x)$,
\begin{equation}
P_{\alpha }\longrightarrow h^{-1}(x)P_{\alpha }h(x),\qquad Q_{\alpha
}\rightarrow h^{-1}(x)Q_{\alpha }h(x)+h^{-1}(x)\partial _{\alpha }h(x)
\label{transf3}
\end{equation}
and all the discussion for the earlier case can again be carried through.

\section{Monodromy Matrix}

In the last section, we saw that the two dimensional string effective
action, dimensionally reduced from $D$-dimensions, has a natural description
of a sigma model defined on a coset, coupled to gravity. In the case when
there are no Abelian gauge fields present in the starting string action, the
sigma model is defined on the coset ${\frac{O(d,d)}{O(d)\times O(d)}}$ where
$d=D-2$. On the other hand, if $n$ Abelian gauge fields are present in the
starting string action, the coset can be identified with ${\frac{O(d,d+n)}{
O(d)\times O(d+n)}}$. In this section, we will
further analyze the integrability properties of such a system and construct
the monodromy matrix associated with the system.

Let us consider a general sigma model in flat space-time, defined on the
coset $G/H$. The two cases of interest for us are when $G=O(d,d),H=O(d)
\times O(d)$ and $G=O(d,d+n),H=O(d)\times O(d+n)$. Let $V\in G/H$ and $
M=VV^{T}$. Then, as we have noted in the last section, we can decompose the
current $V^{-1}\partial _{\alpha }V$ belonging to the Lie algebra of $G$ as
\begin{equation}
V^{-1}\partial _{\alpha }V=P_{\alpha }+Q_{\alpha }  \label{vdv}
\end{equation}
where $Q_{\alpha }$ belongs to the Lie algebra of $H$, while $P_{\alpha }$
belongs to the complement. The integrability condition, following from this,
corresponds to the zero curvature condition
\begin{equation}
\partial _{\alpha }(V^{-1}\partial _{\beta }V)-\partial _{\beta
}(V^{-1}\partial _{\alpha }V)+[(V^{-1}\partial _{\alpha }V),(V^{-1}\partial
_{\beta }V)]=0  \label{intcond}
\end{equation}
Explicitly, this equation gives
\begin{eqnarray}
\partial _{\alpha }Q_{\beta }-\partial _{\beta }Q_{\alpha }+[Q_{\alpha
},Q_{\beta }]+[P_{\alpha },P_{\beta }] &=&0  \nonumber \\
D_{\alpha }P_{\beta }-D_{\beta }P_{\alpha } &=&0  \label{j1}
\end{eqnarray}
where we have defined
\begin{equation}
D_{\alpha }P_{\beta }=\partial _{\alpha }P_{\beta }+[Q_{\alpha },P_{\beta }]
\label{x1}
\end{equation}
The equations of motion for the flat space sigma model (see eq. (\ref{eq3}))
\begin{equation}
\eta ^{\alpha \beta }\partial _{\alpha }(M^{-1}\partial _{\beta }M)=0
\label{j2}
\end{equation}
can be rewritten in the form
\begin{equation}
\eta ^{\alpha \beta }D_{\alpha }P_{\beta }=0  \label{x2}
\end{equation}

Let us next introduce a one parameter family of matrices $\hat{V}(x,t)$
where $t$ is a constant parameter (and not time), also known as the spectral
parameter, such that $\hat{V}(x,t=0)=V(x)$ and
\begin{equation}
\hat{V}^{-1}\partial _{\alpha }\hat{V}=Q_{\alpha }+{\frac{1+t^{2}}{1-t^{2}}}
\,P_{\alpha }+{\frac{2t}{1-t^{2}}}\,\epsilon _{\alpha \beta }P^{\beta }
\label{x3}
\end{equation}
Then, it is straightforward to check that the integrability condition
\begin{equation}
\partial _{\alpha }(\hat{V}^{-1}\partial _{\beta }\hat{V})-\partial _{\beta
}(\hat{V}^{-1}\partial _{\alpha }\hat{V})+[(\hat{V}^{-1}\partial _{\alpha }
\hat{V}),(\hat{V}^{-1}\partial _{\beta }\hat{V})]=0  \label{j3}
\end{equation}
leads naturally to eqs. (\ref{j1}),(\ref{x2}). Namely, the
integrability  condition (\ref{j1}) as well as the equation of motion
for the  flat space sigma model
are obtained from the zero curvature condition associated with a potential
which depends on a constant spectral parameter.

In the presence of gravity, however, the equation for the sigma model
modifies \cite{nic2,nic3}. In the conformal gauge $g_{\alpha \beta
}=e^{\bar{\phi}}\eta _{\alpha \beta }$, eq. (\ref{eq3}) takes the form
\begin{eqnarray}
\eta ^{\alpha \beta }\partial _{\alpha }(e^{-\bar{\phi}}M^{-1}\partial
_{\beta }M) &=&0  \nonumber \\
{\rm or,}\quad \eta ^{\alpha \beta }D_{\alpha }(e^{-\bar{\phi}}P_{\beta
})=D_{\alpha }(e^{-\bar{\phi}}P^{\alpha }) &=&0  \label{x5}
\end{eqnarray}
As before, we can introduce a one parameter family of potentials depending
on a spectral parameter and with a  decomposition of the form
(\ref{x3}).  However, in this case, it is easy to check that the
zero curvature condition in (\ref{j3}) leads to the correct dynamical
equation as well as the integrability condition provided the spectral
parameter is space-time dependent and satisfies
\begin{equation}
\partial _{\alpha }t=-{\frac{1}{2}}\epsilon _{\alpha \beta }\partial ^{\beta
}\left( e^{-\bar{\phi}}(t+{\frac{1}{t}})\right)   \label{j4}
\end{equation}
In the conformal gauge, as we have seen earlier in (\ref{dilaton}), the
shifted dilaton satisfies a simple equation. Therefore, defining
\begin{equation}
\rho (x)=e^{-\bar{\phi}}  \label{j5}
\end{equation}
we note that the solution following from the equation for the shifted
dilaton can be written as
\begin{equation}
\rho (x)=\rho _{+}(x^{+})+\rho _{-}(x^{-})  \label{j6}
\end{equation}
With this, the solution to eq. (\ref{j4}) can be written as
\begin{equation}
t(x)={\frac{\sqrt{\omega +\rho _{+}}-\sqrt{\omega -\rho _{-}}}{\sqrt{\omega
+\rho _{+}}+\sqrt{\omega -\rho _{-}}}}  \label{spectpar}
\end{equation}
where $\omega $ is the constant of integration, which can be thought of as a
global spectral parameter. It is clear that the solutions in eq. (\ref
{spectpar}) are double valued in nature.

There are several things to note from our discussion so far. First of all,
the one parameter family of connections (currents) does not determine the
potential $\hat{V}(x,t)$ uniquely, namely, $\hat{V}$ and $S(\omega )\hat{V}$
, where $S(\omega )$ is a constant matrix, yield the same one parameter
family of connections. Second, in the presence of the spectral parameter,
the symmetric space automorphism can be generalized as
\begin{equation}
\eta ^{\infty }(\hat{V}(x,t))=\eta (\hat{V}(x,{\frac{1}{t}}))=\left( \hat{V}
^{-1}(x,{\frac{1}{t}})\right) ^{T}  \label{x6}
\end{equation}
It can be shown, following from this, that
\begin{equation}
\left( \hat{V}^{-1}(x,{\frac{1}{t}})\partial _{\alpha }\hat{V}(x,{\frac{1}{t}
})\right) ^{T}=-\hat{V}^{-1}(x,t)\partial _{\alpha }\hat{V}(x,t)  \label{x7}
\end{equation}
Given these, let us define
\begin{equation}
{\cal M}=\hat{V}(x,t)\hat{V}^{T}(x,{\frac{1}{t}})  \label{x8}
\end{equation}
It follows now, from eq. (\ref{x7}), that
\begin{equation}
\partial _{\alpha }{\cal M}=0  \label{x9}
\end{equation}
Namely, ${\cal M}={\cal M}(\omega )$ and is independent of space-time
coordinates. ${\cal M}(\omega )$ is known as the monodromy matrix for the
system under study and encodes properties of integrability such as the
conserved quantities associated with the system.

Let us next describe how the monodromy matrix is constructed for such
systems. For simplicity, we will consider the action in (\ref{action2}),
which describes a sigma model defined on the coset ${\frac{O(d,d)}{%
O(d)\times O(d)}}$. The other case can be studied in a completely analogous
manner. To start with, let us set the anti-symmetric tensor field to zero,
namely, $B=0$. In this case, we can write
\begin{equation}
M^{(B=0)}=\left(
\begin{array}{cc}
G^{-1} & 0 \\
0 & G
\end{array}
\right) ,\qquad V^{(B=0)}=\left(
\begin{array}{cc}
E^{-1} & 0 \\
0 & E
\end{array}
\right)  \label{x10}
\end{equation}
Let us further assume that the matrix $E$ and, therefore, $G$ are diagonal,
as is relevant in the study of colliding plane waves. Namely, let us
parameterize

\begin{eqnarray}
E &=&{\rm diag.}\,\left( e^{{\frac{1}{2}}(\lambda +\psi _{1})},e^{{\frac{1}{2
}}(\lambda +\psi _{2})},\cdots ,e^{{\frac{1}{2}}(\lambda +\psi _{d})}\right)
\nonumber \\
G &=&{\rm diag.}\,\left( e^{\lambda +\psi _{1}},e^{\lambda +\psi
_{2}},\cdots ,e^{\lambda +\psi _{d}}\right)  \label{EG}
\end{eqnarray}
with $\sum_{i}\psi _{i}=0$ so that $\lambda ={\frac{1}{d}}\log \det G$, as
adopted in \cite{bv}.

In this case, it follows that (see (\ref{current}))
\begin{eqnarray}
P_{\alpha } &=&{\frac{1}{2}}\left( (V^{(B=0)})^{-1}\partial _{\alpha
}V^{(B=0)}+((V^{(B=0)})^{-1}\partial _{\alpha }V^{(B=0)})^{T}\right) =\left(
\begin{array}{cc}
-E^{-1}\partial _{\alpha }E & 0 \\
0 & E^{-1}\partial _{\alpha }E
\end{array}
\right)  \nonumber \\
Q_{\alpha } &=&{\frac{1}{2}}\left( (V^{(B=0)})^{-1}\partial _{\alpha
}V^{(B=0)}-((V^{(B=0)})^{-1}\partial _{\alpha }V^{(B=0)})^{T}\right) =0
\label{pq3}
\end{eqnarray}
so that we have
\begin{equation}
(\hat{V}^{(B=0)})^{-1}\partial _{+}\hat{V}^{(B=0)}={\frac{1-t}{1+t}}
\,P_{+},\qquad (\hat{V}^{(B=0)})^{-1}\partial _{-}\hat{V}^{(B=0)}={\frac{1+t
}{1-t}}\,P_{-}  \label{vdv2}
\end{equation}

Since $P_{\pm }$ are diagonal matrices and $\hat{V}
^{(B=0)}(x,t=0)=V^{(B=0)}(x)$ is diagonal, it follows that we can write
\begin{equation}
\hat{V}^{(B=0)}(x,t)=\left(
\begin{array}{cc}
\overline{V}^{(B=0)}(x,t) & 0 \\
0 & (\overline{V}^{(B=0)})^{-1}(x,t)
\end{array}
\right)  \label{vb0}
\end{equation}
with $\overline{V}^{(B=0)}(x,t)$ a diagonal matrix of the form $(\overline{V}
_{1},\overline{V}_{2},\cdots ,\overline{V}_{d})$. Let us assume that
\begin{equation}
\overline{V}_{i}={\frac{t_{d+i}}{t_{i}}}\,{\frac{t-t_{i}}{t-t_{d+i}}}
\,E_{i}^{-1},\qquad i=1,2,\cdots ,d  \label{Vi}
\end{equation}
where $t_{i}$ is the spectral parameter corresponding to the constant $
\omega _{i}$. Clearly, for $t=0$, this leads to the diagonal elements of $
V^{(B=0)}$. Furthermore, noting the form of $P_{\pm }$ in (\ref{pq3}) and
recalling that the spectral parameters satisfy
\begin{equation}
\partial _{\pm }t={\frac{1\mp t}{1\pm t}}\,\partial _{\pm }\ln \rho ,\qquad
\partial _{\pm }t_{i}={\frac{1\mp t_{i}}{1\pm t_{i}}}\,\partial _{\pm }\ln
\rho  \label{2rel}
\end{equation}
it is easy to verify that
\begin{equation}
\overline{V}_{i}^{-1}\partial _{\pm }\overline{V}_{i}=\partial _{\pm }\ln
E_{i}^{-1}\mp {\frac{t}{1\pm t}}\partial _{\pm }\ln \left( -{\frac{t_{i}}{
t_{d+i}}}\right) ={\frac{1\mp t}{1\pm t}}\,\partial _{\pm }\ln E_{i}^{-1}
\label{VidVi}
\end{equation}
provided we identify ($t_{i}$ and $t_{d+i}$ have opposite signatures
following from the double valued nature of the solutions in
(\ref{spectpar})) 
\begin{equation}
-{\frac{t_{i}}{t_{d+i}}}=E_{i}^{-2}  \label{connect1}
\end{equation}
In that case, we can write
\begin{equation}
\left( \hat{V}^{(B=0)}\right) ^{-1}\partial _{\pm }\hat{V}^{(B=0)}={\frac{
1\mp t}{1\pm t}}\left(
\begin{array}{cc}
-E^{-1}\partial _{\pm }E & 0 \\
0 & E^{-1}\partial _{\pm }E
\end{array}
\right) ={\frac{1\mp t}{1\pm t}}\,P_{\pm }  \label{true1}
\end{equation}

Thus, we see that, in the present case,
\begin{equation}
\overline{V}_{i}={\frac{t_{d+i}}{t_{i}}}\,{\frac{t-t_{i}}{t-t_{d+i}}}
\,E_{i}^{-1}=\sqrt{-{\frac{t_{d+i}}{t_{i}}}}\,{\frac{t-t_{i}}{t-t_{d+i}}}
\label{finalVi}
\end{equation}
and the matrix $\hat{V}^{(B=0)}(x,t)$ has $2d$ simple poles - one pair for
every diagonal element $E_{i}$. Furthermore, it is simple to check from eq.
( \ref{spectpar}) that the spectral parameters satisfy
\begin{equation}
{\frac{\omega -\omega _{i}}{\omega -\omega _{d+i}}}={\frac{t_{d+i}}{t_{i}}}
\, {\frac{t-t_{i}}{t-t_{d+i}}}\,{\frac{{\frac{1}{t}}-t_{i}}{{\frac{1}{t}}
-t_{d+i}}}  \label{relat2}
\end{equation}
so that we can determine the monodromy matrix to be of the form
\begin{equation}
\widehat{{\cal M}}^{(B=0)}=\hat{V}^{(B=0)}(x,t)(\hat{V}^{B=0)})^{T}(x,{\frac{
1}{t}})=\left(
\begin{array}{cc}
{\cal M}(\omega ) & 0 \\
0 & {\cal M}^{-1}(\omega )
\end{array}
\right)
\end{equation}
where ${\cal M}(\omega )$ is diagonal with
\begin{equation}
{\cal M}_{i}(\omega )=\overline{V}_{i}(x,t)\overline{V}_{i}(x,{\frac{1}{t}}
)=-{\frac{\omega -\omega _{i}}{\omega -\omega _{d+i}}}
\end{equation}
We note that the double valued relation between the global and the local
spectral parameters allows us to choose $\omega _{d+i}=-\omega _{i}$, in
which case, we have
\begin{equation}
{\cal M}_{i}(\omega )={\frac{\omega _{i}-\omega }{\omega _{i}+\omega }}
\end{equation}
This determines the monodromy matrix for the case when $B=0$.

Let us note next that we are dealing with a sigma model, obtained through
dimensional reduction of a higher dimensional tree level string effective
action. Therefore, the symmetries present in the string theory, such
as  $T$~-~duality, should be encoded in the monodromy matrix as
well. For  example, it
is known that one can generate new backgrounds (of the string theory)
starting from given ones through $T$-duality transformations. In particular,
starting from a background where $B=0$, it is possible, in some models (such
as the Nappi-Witten model), to generate backgrounds with $B\neq 0$ through a $
T$-duality rotation. It is natural, therefore, to examine how the monodromy
matrix transforms under such transformations, for, then, we can determine
the monodromy matrix for more complicated backgrounds starting from simpler
ones.

Let us note that the $T$-duality transformation, within the context of
string theory (without Abelian gauge fields), corresponds to a global
$O(d,d)$  rotation. Since the one parameter family of
matrices $\hat{V}(x,t)\in {\frac{O(d,d)}{O(d)\times O(d)}}$ much like $V(x)=
\hat{V}(x,t=0)$, it follows that under a global $O(d,d)$ rotation
\begin{eqnarray}
\hat{V}(x,t) &\longrightarrow &\Omega ^{T}\hat{V}(x,t)  \nonumber \\
\widehat{{\cal M}}(\omega )=\hat{V}(x,t)\hat{V}^{T}(x,{\frac{1}{t}})
&\longrightarrow &\Omega ^{T}\hat{V}(x,t)\hat{V}^{T}(x,{\frac{1}{t}})\Omega
=\Omega ^{T}\widehat{{\cal M}}(\omega )\Omega  \label{WideM}
\end{eqnarray}
Let us also recall that, under a local $O(d)\times O(d)$ transformation, $
\hat{V}(x,t)\rightarrow \hat{V}(x,t)h(x)$. Therefore, the only local
transformations which will preserve the global nature of the monodromy
matrix are the ones that do not depend on the local spectral parameter
explicitly. We have already seen that the matrix $M=VV^{T}$ is only
sensitive to the global $O(d,d)$ transformations even though $V(x)$
transforms non-trivially under a combined global $O(d,d)$ and a local $
O(d)\times O(d)$ transformation. In a similar manner, $\widehat{{\cal M}}
(\omega )$ is only sensitive to the global $O(d,d)$ rotation. We will check
this explicitly in the case of the Nappi-Witten model in the next section.
For the moment, let us note that this brings out an interesting connection
between the integrability properties of the two dimensional string effective
action and its $T$-duality properties, which can be used as a powerful tool
in determining solutions.

For completeness, we record here the transformation properties of $\widehat{
{\cal M}}$ under an infinitesimal $O(d,d)$ transformation. Let us denote
\begin{equation}
\widehat{{\cal M}}=\left(
\begin{array}{cc}
\widehat{{\cal M}}_{11} & \widehat{{\cal M}}_{12} \\
\widehat{{\cal M}}_{21} & \widehat{{\cal M}}_{22}
\end{array}%
\right)   \label{Mhat1}
\end{equation}
where each element represents a $d\times d$ matrix. Infinitesimally, we can
write \cite{jmj}
\begin{equation}
\Omega =\left(
\begin{array}{cc}
1+X & Y \\
Z & 1+W%
\end{array}
\right)   \label{QXZ}
\end{equation}
where the infinitesimal parameters of the transformation satisfy $
Y^{T}=-Y,Z^{T}=-Z$ and $W=-X^{T}$. Under such an infinitesimal
transformation, it follows from (\ref{QXZ}) that
\begin{eqnarray}
\delta \widehat{{\cal M}}_{11} &=&\widehat{{\cal M}}_{11}X+X^{T}\widehat{
{\cal M}}_{11}-Z\widehat{{\cal M}}_{12}+\widehat{{\cal M}}_{12}Z  \nonumber
\\
\delta \widehat{{\cal M}}_{12} &=&\widehat{{\cal M}}_{11}Y+X^{T}\widehat{
{\cal M}}_{12}-Z\widehat{{\cal M}}_{22}-\widehat{{\cal M}}_{12}X^{T}
\nonumber \\
\delta \widehat{{\cal M}}_{21} &=&-Y\widehat{{\cal M}}_{11}-X\widehat{{\cal M
}}_{21}+\widehat{{\cal M}}_{21}X+\widehat{{\cal M}}_{22}Z  \nonumber \\
\delta \widehat{{\cal M}}_{22} &=&\widehat{{\cal M}}_{21}Y-Y\widehat{{\cal M}
}_{12}-X\widehat{{\cal M}}_{22}-\widehat{{\cal M}}_{22}X^{T}
\label{variations}
\end{eqnarray}

\section{Application}

The ideas presented in the earlier section can be applied to various
physical systems. For example, if we are considering collision of
plane-fronted waves, which correspond to massless states of closed strings,
because of the isometries in the problem, this can be described effectively
by a two dimensional theory and all our earlier discussions can be carried
over \cite{dmm}. In this section, we will discuss one other class of physical
phenomena where our results can be explicitly verified and prove quite
useful.

\subsection{The Nappi-Witten Model}

The Nappi-Witten model \cite{nw} is an example of a cosmological
solution  following
from the string theory. Let us note that, to leading order in $\alpha
^{\prime }$, the string tension, there are several solutions to the string
equations following from (\ref{action1}) that constitute exact CFT
backgrounds. One of these solutions, studied by Nappi and Witten,
corresponds to a gauged ${\frac{SL(2,R)}{SO(1,1)}}\times {\frac{SU(2)}{U(1)}}
$ Wess-Zumino-Witten model and describes a closed expanding universe in $3+1$
dimensions. The backgrounds consist of the metric, the dilaton and the
anti-symmetric tensor fields of the forms (Here, we identify $x^{0}=\tau $.)
\begin{eqnarray}
ds^{2} &=&-d\tau ^{2}+dx^{2}+{\frac{1}{1-\cos 2\tau \cos 2x}}(4\cos ^{2}\tau
\cos ^{2}x\,dy^{2}+4\sin ^{2}\tau \sin ^{2}x\,dz^{2})  \nonumber \\
\phi &=&-\frac{1}{2}\log (1-\cos 2\tau \cos 2x)  \nonumber \\
B_{12} &=&-B_{21}= b = \frac{(\cos 2\tau -\cos 2x)}{(1-\cos 2\tau \cos 2x)}
\label{B1}
\end{eqnarray}
Here, we have set an arbitrary constant parameter appearing in the
Nappi-Witten solution to zero for simplicity.

Note that the backgrounds do not depend on two of the coordinates, namely, $%
(y,z)$ and, consequently, following from our earlier discussions, the system
has an $O(2,2)$ symmetry. It is known that these backgrounds can be obtained
from a much simpler background, with a vanishing $B$ field, of the form
\begin{eqnarray}
ds^{2} &=&-d\tau ^{2}+dx^{2}+{\frac{1}{\tan ^{2}\tau }}\,dy^{2}+\tan
^{2}x\,dz^{2}  \nonumber \\
\overline{\phi } &=&-\log (\sin 2\tau \sin 2x).  \label{ShiftDilat2}
\end{eqnarray}
through an $O(2,2)$ rotation. In the language of our earlier discussion, we
note that we can write
\begin{equation}
G^{B=0}=\left(
\begin{array}{cc}
e^{\lambda ^{(0)}+\psi ^{(0)}} & 0 \\
0 & e^{\lambda ^{(0)}-\psi ^{(0)}}
\end{array}
\right) =\left(
\begin{array}{cc}
\xi _{1} & 0 \\
0 & \xi _{2}
\end{array}
\right)   \label{GB0}
\end{equation}
with
\begin{equation}
\xi _{1}=\exp (\lambda ^{0}+\psi ^{0})=\frac{1}{\tan ^{2}\tau },\qquad \xi
_{2}=\exp (\lambda ^{0}-\psi ^{0})=\frac{1}{\tan ^{2}x}  \label{lambd2}
\end{equation}
where the superscript ``$0$'' denotes the vanishing $B$ field. In this case,
therefore, we have
\begin{equation}
M^{(B=0)}=\left(
\begin{array}{cc}
(G^{(B=0)})^{-1} & 0 \\
0 & G^{(B=0)}
\end{array}
\right)
\end{equation}
On the other hand, it is easy to check that we can write
\begin{eqnarray}
G^{(B)} &=&\left(
\begin{array}{cc}
{\frac{2\xi }{1+\xi _{1}\xi _{2}}} & 0 \\
0 & {\frac{2\xi _{2}}{1+\xi _{1}\xi _{2}}}%
\end{array}
\right)   \nonumber \\
B &=&-{\frac{1-\xi _{1}\xi _{2}}{1+\xi _{1}\xi _{2}}}\,\epsilon =b\epsilon
\end{eqnarray}
where $\epsilon $ is the $2\times 2$ anti-symmetric matrix
\begin{equation}
\epsilon =\left(
\begin{array}{cc}
0 & 1 \\
-1 & 0
\end{array}
\right)   \label{epsilon}
\end{equation}
so that
\begin{equation}
M^{(B)}=\left(
\begin{array}{cc}
(G^{(B)})^{-1} & -(G^{(B)})^{-1}B \\
B(G^{(B)})^{-1} & G^{(B)}-B(G^{(B)})^{-1}B
\end{array}
\right)
\end{equation}
It is now a simple matter to check that
\begin{equation}
M^{(B)}=\Omega ^{T}M^{(B=0)}\Omega
\end{equation}
where \cite{gmv}
\begin{equation}
\Omega ={\frac{1}{\sqrt{2}}}\left(
\begin{array}{cc}
I & \epsilon  \\
\epsilon  & I
\end{array}
\right)   \label{Qepsilon}
\end{equation}
belongs to $O(2,2)$. Here, $I$ represents the $2\times 2$ identity matrix
while $\epsilon $ is the $2\times 2$ anti-symmetric matrix defined in (\ref
{epsilon}). This shows that the backgrounds with a nontrivial anti-symmetric
tensor field can be generated from a much simpler background with a
vanishing $B$ field through a global $O(2,2)$ rotation.

It follows from eq. (\ref{GB0}) that we can write
\begin{equation}
E^{(B=0)}=\left(
\begin{array}{cc}
\sqrt{\xi _{1}} & 0 \\
0 & \sqrt{\xi _{2}}
\end{array}
\right)
\end{equation}
so that we have
\begin{equation}
V^{(B=0)}=\left(
\begin{array}{cc}
(E^{(B=0)})^{-1} & 0 \\
0 & E^{(B=0)}
\end{array}
\right)
\end{equation}
and it follows that (see eq. (\ref{pq3}))
\begin{eqnarray}
P_{\pm }^{(B=0)} &=&\left(
\begin{array}{cc}
-(E^{(B=0)})^{-1}\partial _{\pm }E^{(B=0)} & 0 \\
0 & (E^{(B=0)})^{-1}\partial _{\pm }E^{(B=0)}
\end{array}
\right) =\left(
\begin{array}{cccc}
-{\frac{(\partial _{\pm }\xi _{1})}{\xi _{1}}} & 0 & 0 & 0 \\
0 & -{\frac{(\partial _{\pm }\xi _{2})}{\xi _{2}}} & 0 & 0 \\
0 & 0 & {\frac{(\partial _{\pm }\xi _{1})}{\xi _{1}}} & 0 \\
0 & 0 & 0 & {\frac{(\partial _{\pm }\xi _{2})}{\xi _{2}}}
\end{array}
\right)  \nonumber \\
Q_{\pm }^{(B=0)} &=&0
\end{eqnarray}
In this case, therefore, the one parameter family of potentials, $\hat{V}
^{(B=0)}(x,t)$, have to satisfy (since $Q_{\pm }^{(B=0)}=0$)
\begin{equation}
(\hat{V}^{(B=0)})^{-1}(x,t)\partial _{\pm }\hat{V}^{(B=0)}(x,t)={\frac{1\mp t
}{1\pm t}}\,P_{\pm }^{(B=0)}  \label{VbodVbo}
\end{equation}
where $t$ is the space-time dependent spectral parameter.

Following our earlier construction (see (\ref{finalVi})), we can determine
\begin{equation}
\hat{V}^{(B=0)}(x,t)={\rm diag.}\,(\overline{V}_{1},\cdots ,\overline{V}
_{4})={\rm diag.}\,(\scriptstyle{{\sqrt{-{\frac{t_{3}}{t_{1}}}}{\frac{
t-t_{1} }{t-t_{3}}},\sqrt{-{\frac{t_{4}}{t_{2}}}}{\frac{t-t_{2}}{t-t_{4}}},
\sqrt{-{\ \frac{t_{1}}{t_{3}}}}{\frac{t-t_{3}}{t-t_{1}}},\sqrt{-{\frac{t_{2}
}{t_{4}}}}{\ \frac{t-t_{4}}{t-t_{2}}}}})  \label{diagonalV}
\end{equation}
where
\begin{equation}
-{\frac{t_{1}}{t_{3}}}=(E_{1}^{(B=0)})^{-2}={\frac{1}{\xi _{1}}},\qquad -{\
\frac{t_{2}}{t_{4}}}=(E_{2}^{(B=0)})^{-2}={\frac{1}{\xi _{2}}}
\end{equation}
It can be checked explicitly, using the equation satisfied by the spectral
parameter, (\ref{2rel}), that $\hat{V}^{(B=0)}(x,t)$ in (\ref{diagonalV})
does indeed satisfy (\ref{VbodVbo}). The monodromy matrix, in this case,
follows to be
\begin{eqnarray}
\widehat{{\cal M}}^{(B=0)} &=&\left(
\begin{array}{cc}
{\cal M}(\omega ) & 0 \\
0 & {\cal M}^{-1}(\omega )
\end{array}
\right) =\left(
\begin{array}{cccc}
{\frac{\omega -\omega _{1}}{\omega -\omega _{3}}} & 0 & 0 & 0 \\
0 & {\frac{\omega -\omega _{2}}{\omega -\omega _{4}}} & 0 & 0 \\
0 & 0 & {\frac{\omega -\omega _{3}}{\omega -\omega _{1}}} & 0 \\
0 & 0 & 0 & {\frac{\omega -\omega _{4}}{\omega -\omega _{2}}}
\end{array}
\right)  \nonumber \\
&=&\left(
\begin{array}{cccc}
{\frac{\omega _{1}-\omega }{\omega _{1}+\omega }} & 0 & 0 & 0 \\
0 & {\frac{\omega _{2}-\omega }{\omega _{2}+\omega }} & 0 & 0 \\
0 & 0 & {\frac{\omega _{1}+\omega }{\omega _{1}-\omega }} & 0 \\
0 & 0 & 0 & {\frac{\omega _{2}+\omega }{\omega _{2}-\omega }}
\end{array}
\right)  \label{MhatOmega}
\end{eqnarray}
where we have identified $\omega _{3}=-\omega _{1},\omega _{4}=-\omega _{2}$.

When $B\neq 0$, we can similarly obtain
\begin{eqnarray}
E^{(B)} &=&\left(
\begin{array}{cc}
\sqrt{\frac{2\xi _{1}}{1+\xi _{1}\xi _{2}}} & 0 \\
0 & \sqrt{\frac{2\xi _{2}}{1+\xi _{1}\xi _{2}}}
\end{array}
\right)  \nonumber \\
V^{(B)} &=&\left(
\begin{array}{cc}
(E^{(B)})^{-1} & 0 \\
B(E^{(B)})^{-1} & E^{(B)}
\end{array}
\right)
\end{eqnarray}
It follows now that
\begin{eqnarray}
P_{\pm }^{(B)} &=&\left(
\begin{array}{cc}
-(E^{(B)})^{-1}\partial _{\pm }E^{(B)} & -{\frac{1}{2}}(E^{(B)})^{-1}(
\partial _{\pm }B)(E^{(B)})^{-1} \\
{\frac{1}{2}}(E^{(B)})^{-1}(\partial _{\pm }B)(E^{(B)})^{-1} &
(E^{(B)})^{-1}\partial _{\pm }E^{(B)}
\end{array}
\right)  \nonumber \\
Q_{\pm } &=&\left(
\begin{array}{cc}
0 & {\frac{1}{2}}(E^{(B)})^{-1}(\partial _{\pm }B)(E^{(B)})^{-1} \\
{\frac{1}{2}}(E^{(B)})^{-1}(\partial _{\pm }B)(E^{(B)})^{-1} & 0
\end{array}
\right)
\end{eqnarray}
In this case, therefore, $Q_{\pm }\neq 0$ and the one parameter family of
potentials has to satisfy
\begin{equation}
(\hat{V}^{(B)})^{-1}(x,t)\partial _{\pm }\hat{V}^{(B)}(x,t)=Q_{\pm }^{(B)}+{
\frac{1\mp t}{1\pm t}}\,P_{\pm }^{(B)}  \label{relations3}
\end{equation}
It is straightforward to check that
\begin{equation}
V^{(B)}(x)=\Omega ^{T}V^{(B=0)}(x)h(x)
\end{equation}
where $\Omega $ is the $O(2,2)$ matrix defined in (\ref{Qepsilon}) and $
h(x)\in O(2)\times O(2)$ and is of the form
\begin{equation}
h(x)={\frac{1}{\sqrt{2}}}\left(
\begin{array}{cc}
\sqrt{1-b}I & \sqrt{1+b}\epsilon \\
\sqrt{1+b}\epsilon & \sqrt{1-b}I
\end{array}
\right)
\end{equation}
It can also be checked that
\begin{equation}
\hat{V}^{(B)}(x,t)=\Omega ^{T}\hat{V}^{(B=0)}(x,t)h(x)
\end{equation}
\begin{equation}
={\frac{1}{2}}\left(
\begin{array}{cccc}
\alpha\overline{V}_{1}+\beta\overline{V}_{4} & 0 & 0 & \beta
\overline{V}_{1}-\alpha\overline{V}_{4} \\
0 & \alpha\overline{V}_{2}+\beta\overline{V}_{3} & -\beta
\overline{V}_{2}+\alpha\overline{V}_{3} & 0 \\
0 & -\alpha\overline{V}_{2}+\beta\overline{V}_{3} & \beta
\overline{V}_{2}+\alpha\overline{V}_{3} & 0 \\
\alpha\overline{V}_{1}-\beta\overline{V}_{4} & 0 & 0 & \beta
\overline{V}_{1}+\alpha\overline{V}_{4}
\end{array}
\right)
\end{equation}
where $\alpha=\sqrt{1-b}$ and $\beta=\sqrt{1+b}$, satisfies the defining relation in (\ref{relations3}). 
We note here that,when $B\neq 0$, while $V^{(B)}(x)$ is triangular, $\hat{V}^{(B)}(x,t)$ is
not in general and that both $V^{(B)}$ and $\hat{V}^{(B)}$ are related to
their counterparts with $B=0$ through a combined global $O(2,2)$ and a local
$O(2)\times O(2)$ transformation. Furthermore, as was pointed out earlier,
the local transformation does not depend explicitly on the spectral
parameter. This is quite crucial, for it immediately leads to
\begin{eqnarray}
\widehat{{\cal M}}^{(B)} &=&\hat{V}^{(B)}(x,t)(\hat{V}^{(B)})^{T}(x,{\frac{1
}{t}})=\Omega ^{T}\hat{V}^{(B=0)}(x,t)h(x)h^{T}(x)(\hat{V}^{(B=0)})^{T}(x,{
\frac{1}{t}})\Omega =\Omega ^{T}\widehat{{\cal M}}^{(B=0)}\Omega  \nonumber
\\
&=&{\frac{1}{2}}\left(
\begin{array}{cccc}
{\cal M}_{1}+{\cal M}_{2}^{-1} & 0 & 0 & {\cal M}_{1}-{\cal M}_{2}^{-1} \\
0 & {\cal M}_{2}+{\cal M}_{1}^{-1} & -{\cal M}_{2}+{\cal M}_{1}^{-1} & 0 \\
0 & {\cal M}_{2}+{\cal M}_{1}^{-1} & {\cal M}_{2}+{\cal M}_{1}^{-1} & 0 \\
{\cal M}_{1}-{\cal M}_{2}^{-1} & 0 & 0 & {\cal M}_{1}+{\cal M}_{2}^{-1}
\end{array}
\right)
\end{eqnarray}
where ${\cal M}_{1}={\frac{\omega _{1}-\omega }{\omega _{1}+\omega }}$ and $
{\cal M}_{2}={\frac{\omega _{2}-\omega }{\omega _{2}+\omega }}$ are the two
diagonal elements of ${\cal M}(\omega )$ in (\ref{MhatOmega}). This shows
explicitly that, for backgrounds related by a duality transformation (in the
present example, an $O(2,2)$ rotation), the corresponding monodromy matrices
are also related in a simple manner, as was pointed out in the last section.

\section{Summary and Discussion}

In this talk, we have described the  prescriptions for the construction of the
monodromy  matrix
for two dimensional string effective action. We adopted the procedure 
commonly followed in the construction of the monodromy matrix
for a class of two dimensional $\sigma$-models  in  curved space. As
mentioned earlier, in most of the cases, the $\sigma$-model arises from
the dimensional reduction of higher dimensional Einstein-Hilbert action 
to two dimensional space-time due to the presence of isometries. In the
context of string theory, a similar approach was adopted in the past to
construct the monodromy matrix as was the case with dimensionally
reduced models in gravity.

One of our principal objective was to take into account the symmetries
associated with the string effective action and construct the monodromy
matrix which contains information about these symmetries. We have
succeeded in introducing a procedure for the construction of the monodromy
matrix under general grounds with some mild requirements such as
factorizability and presence of isolated poles. Furthermore, we have
demonstrated that the monodromy matrix transforms non-trivially under 
the non-compact T-duality group when the two dimensional string effective
action respects that symmetry. We feel, this is an interesting and important
result. The procedure, adopted by us, allows us to construct the monodromy
matrix, once a set of string background configurations are known. 
As a result, if we know monodromy matrix for a given set of simple string
vacuum backgrounds, we can directly obtain the corresponding monodromy
matrix for another set of more complicated backgrounds, if the latter
can be  derived by duality
transformations from the simpler backgrounds.

We have discussed an illustrative example in Section {\bf IV} as an
application
of our methods. We considered the Nappi-Witten model which is
exactly solvable for both vanishing and non-vanishing two form potential $B$.
This is a good testing ground for duality transformation properties of
the monodromy matrix. We have constructed this matrix for the case
$B=0$. Subsequently, 
we have also constructed it for the case $B\ne 0$. Then,
as a consistency check, we have derived the ${\widehat {\cal M}}^B$
 from ${\widehat{\cal M}}^{B=0}$ 
following our rules of the transformations of ${\widehat {\cal M}}$
 under duality. Indeed, it is found that the monodromy matrix computed
using the two different ways, mentioned above, coincide. Although we
have not discussed, the
black hole solutions in heterotic string theory can also be described
in this way \cite{dmm2}. Namely,
the  monodromy matrix for
the `seed' Schwarzschild black hole in heterotic string theory can be
constructed and that for charged black hole solutions can be obtained
from this, since
the T-duality transformations which generate charged black hole
solutions  are already
known. 

It is worth while to mention that all our results are derived for the
case of classical two dimensional effective theory as is the case
for effective two dimensional theories derived from higher dimensional
Einstein-Hilbert action. It might be interesting to explore systematically
the construction of the monodromy matrix and its properties in quantum
theory. We hope the work presented here will find applications in diverse
directions where one encounters effective two dimensional models in
the context of string theory. 
 
\vskip .7cm \noindent {\bf Acknowledgment:} One of us (AD) would like
to thank the organizers of the workshop on integrable models for the
kind hospitality. This work is supported in
part by US DOE Grant No. DE-FG 02-91ER40685.

\end{document}